\newif\ifACM
\ACMtrue
\newif\ifIEEE
\newif\ifnobrand

\ifACM
\documentclass[sigconf,9pt,authorversion,nonacm]{acmart}
\author{G. Higgins Hutchinson, E. Sifferman, T. Bhattacharya, \& D. B. Strukov}
\affiliation{\institution{UC Santa Barbara, Department of Electrical and Computer Engineering}\city{Santa Barbara}\state{California}\country{USA}}
\email{{hutch,ethanjsifferman,tinish,dimastrukov}@ucsb.edu}

\acmConference[DAC'24]{Design Automation Conference}{June 2024}{San Francisco, California, USA}

\fi

\ifnobrand
\documentclass[10pt,letterpaper]{article}
\usepackage{hyperref}
\usepackage{natbib}
\usepackage[margin=1in]{geometry}
\author{G. Higgins Hutchinson, E. Sifferman, T. Bhattacharya, \& D. B. Strukov\\UC Santa Barbara, Department of Electrical and Computer Engineering\\\{hutch,ethanjsifferman,tinish,dimastrukov\}@ucsb.edu}
\fi

\ifIEEE
\documentclass[10pt,conference]{IEEEtran}
\usepackage{hyperref}
\newcommand{\citep}{\cite}
\author{
    redacted for anonymous review
}
\fi

\usepackage{graphicx}

\usepackage{gincltex}
\usepackage{physics}
\usepackage{siunitx}
\usepackage{mathtools}

\usepackage{import}

\usepackage{booktabs}

\usepackage{circuitikz}
\usepackage{tikz}
\usetikzlibrary{calc,patterns}
\usetikzlibrary{positioning, shapes.geometric}
\usetikzlibrary{backgrounds}
\usetikzlibrary{shapes.arrows, fadings}
\usetikzlibrary{decorations.markings}
\usetikzlibrary{fit}

\title{
    FPIA: Field-Programmable Ising Arrays with In-Memory Computing
}

\date{November 2023}

\begin{document}

\ifIEEE
\maketitle
\fi

\begin{abstract}
     Ising Machine is a promising computing approach for solving combinatorial optimization problems. It is naturally suited for energy-saving and compact in-memory computing implementations with emerging memories. A na\"ive in-memory computing implementation of quadratic Ising Machine requires an array of coupling weights that grows quadratically with problem size. However, the resources in such an approach are used inefficiently due to sparsity in practical optimization problems. We first show that this issue can be addressed by partitioning a coupling array into smaller sub-arrays. This technique, however, requires interconnecting sub-arrays; hence, we developed in-memory computing architecture for quadratic Ising Machines inspired by island-type field programmable gate arrays, which is the main contribution of our paper. We adapt open-source tools to optimize problem embedding and model routing overhead. Modeling results of benchmark problems for the developed architecture show up to 60x area improvement and faster operation than the baseline approach. Finally, we discuss algorithm/circuit co-design techniques for further improvements.
\end{abstract}

\ifACM
\maketitle
\fi

\section{Introduction}

\begin{figure}
    \begin{tikzpicture}[
        neuron/.style  = {fill=blue!20, regular polygon, regular polygon sides=3, shape border rotate=-90, draw, thick},
        xp/.style = {draw, thick, circle, fill=green!40},
        xplabel/.style = {below left,font=\footnotesize},
        yscale=0.52, xscale=1.0
    ]
    \begin{scope}[xscale=0.9]
        \draw (0,0) node [xp] (xp11) {} node[xplabel] () {$w^{(2)}_{1,1}$};
        \draw (1,0) node [xp] (xp12) {} node[xplabel] () {$w^{(2)}_{1,2}$};
        \draw (3,0) node [xp] (xp1N) {} node[xplabel] () {$w^{(2)}_{1,N}$};
        \draw (0,-1) node [xp] (xp21) {} node[xplabel] () {$w^{(2)}_{2,1}$};
        \draw (1,-1) node [xp] (xp22) {} node[xplabel] () {$w^{(2)}_{2,2}$};
        \draw (3,-1) node [xp] (xp2N) {} node[xplabel] () {$w^{(2)}_{2,N}$};
        \draw (0,-3) node [xp] (xpN1) {} node[xplabel] () {$w^{(2)}_{N,1}$};
        \draw (1,-3) node [xp] (xpN2) {} node[xplabel] () {$w^{(2)}_{N,2}$};
        \draw (3,-3) node [xp] (xpNN) {} node[xplabel] () {$w^{(2)}_{N,N}$};
        \draw (2,-2) node [anchor=center] () {$\ddots$};
        \draw (2,-0.5) node [anchor=center] () {$\ldots$};
        \draw (0.5,-2) node [anchor=center] () {$\vdots$};
        \draw (4,0) node [xp] (xpb1) {} node [xplabel] () {$w^{(1)}_1$};
        \draw (4,-1) node [xp] (xpb2) {} node [xplabel] () {$w^{(1)}_2$};
        \draw (4,-3) node [xp] (xpbN) {} node [xplabel] () {$w^{(1)}_N$};
    
        \draw (5,0) node[neuron] (N1) {} node[above right] () {$x_1$};
        \draw (5,-1) node[neuron] (N2) {} node[above right] () {$x_2$};
        \draw (5,-3) node[neuron] (NN) {} node[above right] () {$x_N$};
    
        \draw (5,-1.8) node [anchor=center] () {$\vdots$};
        \begin{scope}[on background layer]
            \draw (xp11.center) -- (N1);
            \draw (xp21.center) -- (N2);
            \draw (xpN1.center) -- (NN);
            \draw (xpN1.center) -- (xp11.center);
            \draw (xpN2.center) -- (xp12.center);
            \draw (xpNN.center) -- (xp1N.center);
            \draw (N1) -- ++(1.33,0) |- ($(xpN1.center)+(0,-2.0)$) -- (xpN1.center);
            \draw (N2) -- ++(1.0,0) |- ($(xpN2.center)+(0,-1.5)$) -- (xpN2.center);
            \draw (NN) -- ++(0.33,0) |- ($(xpNN.center)+(0,-1.0)$) -- (xpNN.center);
            \draw (xpbN.center) -- ($(xpb1.center)+(0,0.4)$) node [above] () {BIAS};
        \end{scope}
    \end{scope}

    \matrix [draw,anchor=west] at ($(current bounding box.east)+(0.3,-0.1)$) {
        \node [xp,label=right:Coupler] {}; \\
        \node [neuron,label=right:Spin] {}; \\
    };
\end{tikzpicture}
    \vspace{-0.5em}
    \caption{Quadratic Ising Machine for solving $\frac{1}{2}\sum_{i,j}^N W^{(2)}_{ij} x_i x_j + \sum_i^N W^{(1)}_i x_i + W^{(0)}$ QUBO problem. IMs are similar to Hopfield Neural Networks in which ``coupler'' and ``spin'' are called ``synapse'' and ``neuron'', respectively.
    \vspace{-1em}
    }
    \label{fig:hohnn_schematic}
\end{figure}
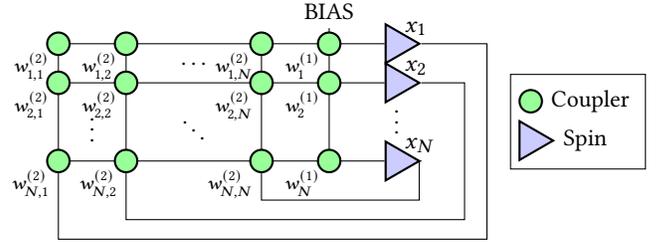

Many practical combinatorial optimization (CO) problems may be represented as Quadratic Unconstrained Binary Optimization (QUBO) \citep{lucasIsingFormulationsMany2014,gloverQuantumBridgeAnalytics2019}. Various physics-based and -inspired platforms have attracted recent interest in building QUBO-solving accelerators --- the majority of such systems are collectively known as quadratic Ising Machines (IM) due to the QUBO merit function's similarity to Lenz' and Ising's Hamiltonian for spin glass systems \citep{lucasIsingFormulationsMany2014}. In the most general, fully connected IMs, the focus of this paper, $N$ spins corresponding to variables in the QUBO function 
are coupled via $O(N^2)$ matrix of tunable couplers that moderate local interaction among pairs of variables (Fig.~\ref{fig:hohnn_schematic}).  Solving a CO problem involves initializing IM states and then updating spin values over time, e.g., at discrete time steps in discrete-time IMs. A new spin state is computed by applying nonlinear (e.g., step) function over a dot product between the current IM state and the spin's coupling weights. With the help of annealing techniques, often implemented in the spin circuitry, IM states can evolve to a ground state, representing a solution to the underlying QUBO problem. 
IM extensions include high-order IMs with coupling among more than two spins \citep{bybee23}, high-dimensional IMs with spin taking on more than two values \citep{strinati22}, and IMs with probabilistic spin updates \citep{dutta21}. The functionality of such more general IMs is similar to other computing approaches for solving CO problems, such as Hopfield Neural Network \citep{hopfieldNeural1982}, Boltzmann machines \citep{sejnowskiHigherOrderBoltzmannMachines1987}, and p-bit computing circuits \citep{camsariStochasticPbitsInvertible2017}.

Optical, electronic, quantum, and hybrid device technologies have been explored for implementing hardware accelerators of IMs and related concepts \citep{mohseni2022}. Because the most common IM operation is a vector-by-matrix multiplication (VMM), In-Memory Computing (IMC) circuit implementations are especially promising. In this work, we will assume that spins are realized with CMOS circuits, while coupling weights are implemented with crossbar-integrated emerging or conventional memory arrays, similar to the previous work on neuromorphic inference accelerators \citep{Bavandpour18}.  

Larger CO problems with $N>1000$ are typically of practical interest. However, the memory array dimensions in IMC circuits are limited, primarily due to IR drop issues \citep{agarwalResistiveMemoryDevice2016}. A solution to this problem in neuromorphic accelerators was to break up a larger single ``logical'' crossbar into multiple smaller size physical crossbar circuits \citep{shafieeISAAC2016, bavandpourACortexEnergyEfficientMultipurpose2020}.  For example, in ISAAC architecture, a tile hosting a smaller physical memory array generates analog partial dot products. Partial products are sent via shared interconnect for the final accumulation performed in the digital domain\citep{shafieeISAAC2016}.  

\begin{figure}
    \begin{adjustbox}{width=3.0in}
    \hspace{-4ex}\import{fig/qubo_sparsity}{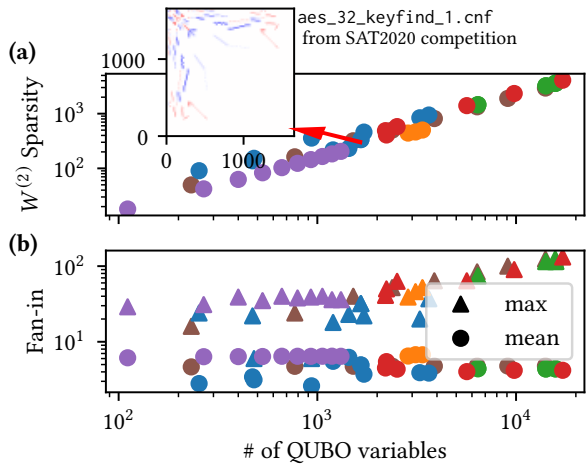}
    \end{adjustbox}
    \caption{(a) Sparsity of the studied benchmark problems (see Fig. \ref{tbl:weight_compression} legend), defined as $N^2$ / (\# nonzero weights). The inset shows a down-sampled coupling matrix with color-coded weight density for one of the problems. (b) Maximum and average spin fan-in (number of incoming non-trivial connections to a spin) for the same problems.  
    }
    \label{fig:factoring_weights}
\end{figure}

On the other hand, practical QUBO-formulated problems are also very sparse (Fig. ~\ref{fig:factoring_weights}a), with the sparsity increasing with the problem size. (The studied benchmark problems are competition 3SAT \citep{SAT2017SolverBenchmarkDescriptions,SAT2020SolverBenchmarkDescriptions}, random uniform 3SAT \citep{SATLIB}, and custom-generated semiprime factoring \citep{PurdomSabryFactoringSAT}, converted to corresponding QUBO problems using Rosenberg approach \citep{lucasIsingFormulationsMany2014}.) In addition, the maximum spin ``fan-in'' (i.e., the maximum number of non-trivial elements in a single row of a coupling matrix) is much smaller than $N$ (Fig. ~\ref{fig:factoring_weights}b). Hence, all nontrivial coupling weights of a spin might be implemented within a single physical crossbar, avoiding the need for expensive ADC circuits.  In fact, sparseness and limited fan-in are necessary attributes for hard (most relevant) decision-type SAT problems. For example, the clause-to-variable ratio for hard 3SAT problems is close to 4.5, translating to the linear sparseness scaling with the size of equivalent QUBO problems.

These observations motivate our work --- to develop more advanced Field-Programmable Ising Machine (``FPIA'') architecture based on efficient IMC circuits and relevant design automation algorithms that take advantage of sparsity and limited spin fan-in of practical QUBO problems. Furthermore, we focus on high-performance solutions that could exploit massively spin update parallelism and rapid convergence of IMs \citep{hizzaniMemristorBasedPUBO2023}. While there have been prior reports on IMC-enabled emerging memory IM architectures\citep{ipek16,sharmaIsingMachineMultiChip}, and, separately, on reconfigurable spintronics-based IMs\citep{mondalIsingFPGASpintronicsbasedReconfigurable2021}, we believe that our work is the first to report programmable IM architecture that efficiently integrates IMC and rich FPGA-like interconnect to optimally implement locally dense, globally sparse connectivity of practical QUBO problems.     

\section{Weight Utilization Improvement}
To make FPIA's motivation more concrete, we investigate the prospects of partitioning problems' weight arrays into smaller sub-arrays such that all non-zero weights of each spin are located in the same sub-array. The key feature to exploit in a packing algorithm is the flexibility in mapping QUBO variables to IM spins, i.e., a QUBO variable can be mapped to any hardware spin without affecting IM's functionality. We use a quadratic-time greedy algorithm that is similar to the well-known First-Fit-Decreasing (FFD) algorithm for integer bin-packing algorithm because performing packing optimally is known to be NP-hard \citep{izumiComputationalComplexityPacking1998}. The algorithm starts by creating a list of (unpacked) spins sorted by their fan-in and a list of initially empty sub-arrays representing the clusters. Each spin steps in order through the list of clusters, and is packed into the first valid cluster. A candidate spin may be invalid to add to a cluster either because the cluster already contains $O$ total spins, or because the cardinality of the union of spin inputs after adding the candidate would exceed $I$. We test this approximate packing for several QUBOs, holding $I=256$ and varying $O$.

The results show that partitioning significantly improves weight utilization, especially for larger QUBO problems because of larger available sparsity (Fig.~\ref{fig:factoring_weights}a). As expected, it is the largest for $O=1$; however, this is not necessarily representative of higher circuit density because such an analysis only includes memory array area but neglects array periphery and routing overheads, which would be higher for smaller sub-arrays. We next introduce the FPIA architecture and perform its detailed modeling to understand these tradeoffs better. 

\begin{figure}
         \import{fig/greedy_packing}{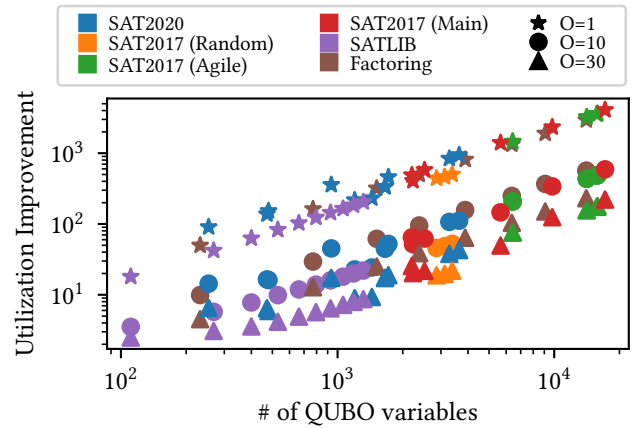}
    \caption{
    Coupling weight utilization improvement by tiling, defined as $N^2 / (\sum_j I_j \cdot O_j$), where $I_j \leq I$ and $O_j \leq O$ are the used input and output sizes of a cluster, and $j$ steps through all clusters created by the FFD algorithm.
    }
    \label{tbl:weight_compression}
\end{figure}

\section{FPIA and Baseline Architectures}

\begin{figure}[h]
    \centering
    \includegraphics[width = 180 pt]{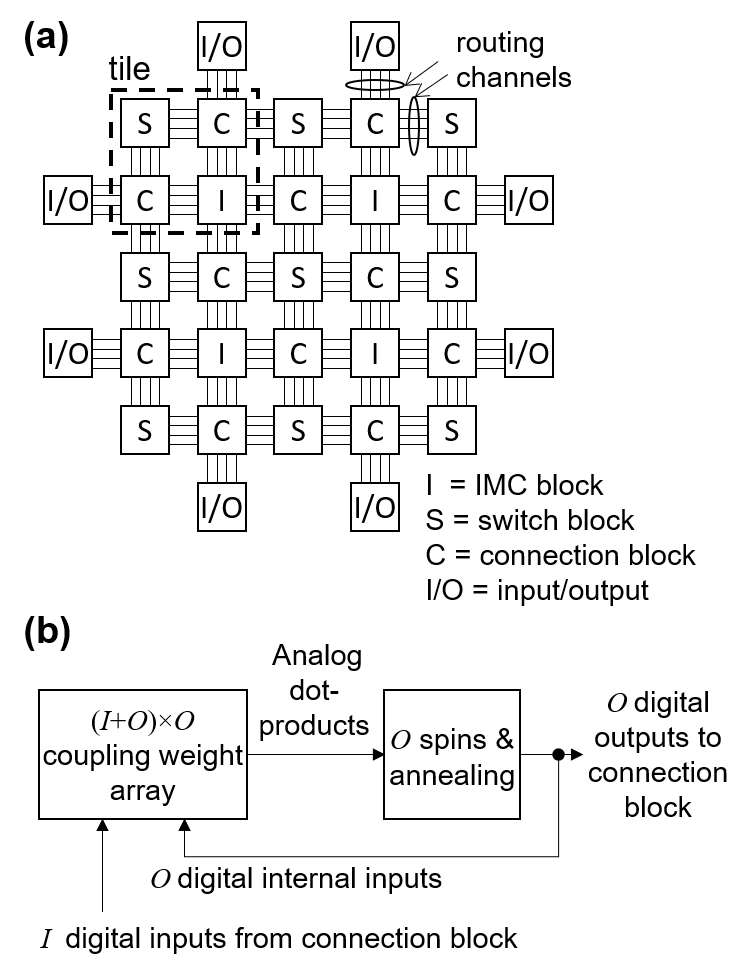}
    \caption{(a) FPIA's top-level and (b) mixed-signal IMC block architectures. Only four tiles are shown in (a) for simplicity.}
    \label{fig:fpia_arch}
\end{figure}

The proposed FPIA architecture resembles an island-type FPGA \citep{betzArchCAD1999_2}, with the configurable logic blocks (CLBs) replaced by mixed-signal IMC blocks (Fig. \ref{fig:fpia_arch}).  IMC block's $I$ input and $O$ output pins are connected to the routing channels via the nearest ``connection block'', but only to a fraction of wires (denoted with $F_\textrm{I}$ and $F_\textrm{O}$, respectively) as is common in island-type FPGAs. Routing channels contain wires stretching $R_{\textrm{tile}}$ tiles vertically or horizontally. Bends and extensions of the wire are implemented with the ``switch blocks''. 

At the core of the IMC block is a $(I+O)\times O$ crossbar array of multi-bit memory cells and associated peripheral circuitry for implementing up to $O$ spins, most importantly including two-quadrant (i.e., differential pair to encode negative weights) mixed-signal VMM, nonlinear activation, and annealing, and all spins' coupling weights. The digital inputs to the crossbar could be local, i.e., routed inside the block from local spins, or supplied externally from spins of other IMC blocks (Fig. \ref{fig:fpia_arch}b). Such an architecture allows up to $I$ global and up to $O$ local couplings implemented for each spin (though a lesser degree of coupling is assumed in modeling experiments, as discussed in section 4).

\begin{figure}
    \centering
    \includegraphics[width=200 pt]{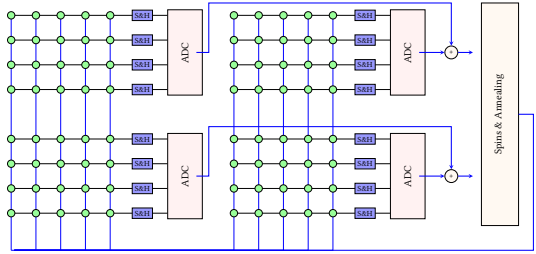}
    \caption{Baseline architecture. }
    \label{fig:isaac_like_arch}
\end{figure}

To provide a fair point of comparison for FPIA, we consider a baseline architecture inspired by ISAAC \citep{shafieeISAAC2016} in which a full logical coupling matrix is again partitioned into smaller ($S\times S$) physical sub-arrays but without previously considered optimization related to matrix sparsity (Fig.~\ref{fig:isaac_like_arch}). The upside of such straightforward (``naive'') implementation is very simple routing. As discussed in the introduction section, its downside is the need to add partial products from multiple sub-arrays to compute spin states. Similarly to FPIA, we assume IMC blocks with digital outputs. Block's partial dot-products are digitized, and top-level aggregation of results is performed in the digital domain. To further simply the routing and reduce ADC overhead, similar to ISAAC \citep{shafieeISAAC2016}, we assume that IMC computations are run on a (relatively) slow clock, then sampled-and-held while a fast, pipelined shared ADC digitizes many intermediate values in a single IMC clock cycle.        

\section{Modeling Framework}

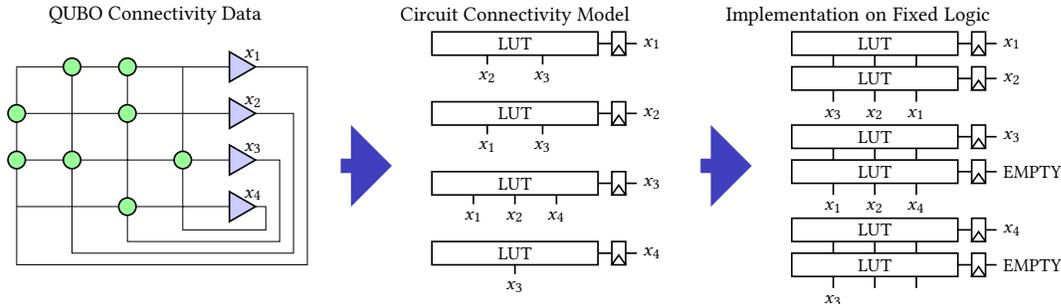
\begin{figure*}
    \centering
    \vspace{-0.5em}
    \begin{adjustbox}{width=0.85\textwidth}
    \begin{tikzpicture}[
        neuron/.style  = {draw,thick,fill=blue!20, regular polygon, regular polygon sides=3, shape border rotate=-90},
        xp/.style = {draw,thick,circle, fill=green!40},
        sh/.style = {draw,thick,rectangle, anchor=center, fill=blue!40},
        xscale=0.95,yscale=0.8
    ]
    
    \begin{scope}[xshift=0,yshift=0]
        \draw (0,-1) node [xp] (xp01) {};
        \draw (0,-2) node [xp] (xp02) {};
        \draw (1,-0) node [xp] (xp10) {};
        \draw (1,-2) node [xp] (xp12) {};
        \draw (2,-0) node [xp] (xp20) {};
        \draw (2,-1) node [xp] (xp21) {};
        \draw (2,-3) node [xp] (xp23) {};
        \draw (3,-2) node [xp] (xp32) {};
        \draw (4,-0) node [neuron] (n0) {} node [above right] () {$x_1$};
        \draw (4,-1) node [neuron] (n1) {} node [above right] () {$x_2$};
        \draw (4,-2) node [neuron] (n2) {} node [above right] () {$x_3$};
        \draw (4,-3) node [neuron] (n3) {} node [above right] () {$x_4$};
        \begin{scope}[on background layer]
            \draw (0,-0) -- (5.25,-0) -- (5.25,-4.25) -- (0,-4.25) -- (0,-0);
            \draw (0,-1) -- (5.0,-1) -- (5.0,-4.0) -- (1,-4.0) -- (1,-0);
            \draw (0,-2) -- (4.75,-2) -- (4.75,-3.75) -- (2,-3.75) -- (2,-0);
            \draw (0,-3) -- (4.5,-3) -- (4.5,-3.5) -- (3,-3.5) -- (3,-0);
        \end{scope}
    \end{scope}

    \draw (6.25,-2.125) node [single arrow, fill=blue!75!yellow, single arrow head extend=15.0, anchor=center, minimum width=40, minimum height=25] () {};

    \begin{scope}[xshift=7cm,yshift=1.0cm]
        \draw [draw, thick] (0.5,-0.25) rectangle node [] () {LUT} (3.5,-0.75);
        \draw [thick] (1.5,-0.75) -- +(0,-0.2) node [below] () {$x_2$};
        \draw [thick] (2.5,-0.75) -- +(0,-0.2) node [below] () {$x_3$};
        \draw [thick] (3.75,-0.75) -- (3.75,-0.25) -- (4.0,-0.25) -- (4.0,-0.75) -- (3.75,-0.75) -- (3.875,-0.54) -- (4.0,-0.75);
        \draw [thick] (3.5,-0.5) -- (3.75,-0.5);
        \draw [thick] (4.0,-0.5) -- +(0.2,0.0) node [right] () {$x_1$};
    \end{scope}
    \begin{scope}[xshift=7cm,yshift=-0.5cm]
        \draw [draw, thick] (0.5,-0.25) rectangle node [] () {LUT} (3.5,-0.75);
        \draw [thick] (1.5,-0.75) -- +(0,-0.2) node [below] () {$x_1$};
        \draw [thick] (2.5,-0.75) -- +(0,-0.2) node [below] () {$x_3$};
        \draw [thick] (3.75,-0.75) -- (3.75,-0.25) -- (4.0,-0.25) -- (4.0,-0.75) -- (3.75,-0.75) -- (3.875,-0.54) -- (4.0,-0.75);
        \draw [thick] (3.5,-0.5) -- (3.75,-0.5);
        \draw [thick] (4.0,-0.5) -- +(0.2,0.0) node [right] () {$x_2$};
    \end{scope}
    \begin{scope}[xshift=7cm,yshift=-2cm]
        \draw [draw, thick] (0.5,-0.25) rectangle node [] () {LUT} (3.5,-0.75);
        \draw [thick] (1.25,-0.75) -- +(0,-0.2) node [below] () {$x_1$};
        \draw [thick] (2.0,-0.75) -- +(0,-0.2) node [below] () {$x_2$};
        \draw [thick] (2.75,-0.75) -- +(0,-0.2) node [below] () {$x_4$};
        \draw [thick] (3.75,-0.75) -- (3.75,-0.25) -- (4.0,-0.25) -- (4.0,-0.75) -- (3.75,-0.75) -- (3.875,-0.54) -- (4.0,-0.75);
        \draw [thick] (3.5,-0.5) -- (3.75,-0.5);
        \draw [thick] (4.0,-0.5) -- +(0.2,0.0) node [right] () {$x_3$};
    \end{scope}
    \begin{scope}[xshift=7cm,yshift=-3.5cm]
        \draw [draw, thick] (0.5,-0.25) rectangle node [] () {LUT} (3.5,-0.75);
        \draw [thick] (2.0,-0.75) -- +(0,-0.2) node [below] () {$x_3$};
        \draw [thick] (3.75,-0.75) -- (3.75,-0.25) -- (4.0,-0.25) -- (4.0,-0.75) -- (3.75,-0.75) -- (3.875,-0.54) -- (4.0,-0.75);
        \draw [thick] (3.5,-0.5) -- (3.75,-0.5);
        \draw [thick] (4.0,-0.5) -- +(0.2,0.0) node [right] () {$x_4$};
    \end{scope}

    \draw (12.75,-2.125) node [single arrow, fill=blue!75!yellow, single arrow head extend=15.0, anchor=center, minimum width=40, minimum height=25] () {};

    \begin{scope}[xshift=13.5cm,yshift=1cm]
        \begin{scope}[yshift=0cm]
            \draw [draw, thick] (0.5,-0.25) rectangle node [] () {LUT} (3.5,-0.75);
            \draw [thick] (1.25,-0.75) -- +(0,-0.25);
            \draw [thick] (2.0,-0.75) -- +(0,-0.25);
            \draw [thick] (2.75,-0.75) -- +(0,-0.25);
            \draw [thick] (3.75,-0.75) -- (3.75,-0.25) -- (4.0,-0.25) -- (4.0,-0.75) -- (3.75,-0.75) -- (3.875,-0.54) -- (4.0,-0.75);
            \draw [thick] (3.5,-0.5) -- (3.75,-0.5);
            \draw [thick] (4.0,-0.5) -- +(0.2,0.0) node [right] () {$x_1$};
        \end{scope}
        \begin{scope}[yshift=-0.75cm]
            \draw [draw, thick] (0.5,-0.25) rectangle node [] () {LUT} (3.5,-0.75);
            \draw [thick] (1.25,-0.75) -- +(0,-0.25) node [below] () {$x_3$};
            \draw [thick] (2.0,-0.75) -- +(0,-0.25) node [below] () {$x_2$};
            \draw [thick] (2.75,-0.75) -- +(0,-0.25) node [below] () {$x_1$};
            \draw [thick] (3.75,-0.75) -- (3.75,-0.25) -- (4.0,-0.25) -- (4.0,-0.75) -- (3.75,-0.75) -- (3.875,-0.54) -- (4.0,-0.75);
            \draw [thick] (3.5,-0.5) -- (3.75,-0.5);
            \draw [thick] (4.0,-0.5) -- +(0.2,0.0) node [right] () {$x_2$};
        \end{scope}
    \end{scope}

    \begin{scope}[xshift=13.5cm,yshift=-1cm]
        \begin{scope}[yshift=0cm]
            \draw [draw, thick] (0.5,-0.25) rectangle node [] () {LUT} (3.5,-0.75);
            \draw [thick] (1.25,-0.75) -- +(0,-0.25);
            \draw [thick] (2.0,-0.75) -- +(0,-0.25);
            \draw [thick] (2.75,-0.75) -- +(0,-0.25);
            \draw [thick] (3.75,-0.75) -- (3.75,-0.25) -- (4.0,-0.25) -- (4.0,-0.75) -- (3.75,-0.75) -- (3.875,-0.54) -- (4.0,-0.75);
            \draw [thick] (3.5,-0.5) -- (3.75,-0.5);
            \draw [thick] (4.0,-0.5) -- +(0.2,0.0) node [right] () {$x_3$};
        \end{scope}
        \begin{scope}[yshift=-0.75cm]
            \draw [draw, thick] (0.5,-0.25) rectangle node [] () {LUT} (3.5,-0.75);
            \draw [thick] (1.25,-0.75) -- +(0,-0.25) node [below] () {$x_1$};
            \draw [thick] (2.0,-0.75) -- +(0,-0.25) node [below] () {$x_2$};
            \draw [thick] (2.75,-0.75) -- +(0,-0.25) node [below] () {$x_4$};
            \draw [thick] (3.75,-0.75) -- (3.75,-0.25) -- (4.0,-0.25) -- (4.0,-0.75) -- (3.75,-0.75) -- (3.875,-0.54) -- (4.0,-0.75);
            \draw [thick] (3.5,-0.5) -- (3.75,-0.5);
            \draw [thick] (4.0,-0.5) -- +(0.2,0.0) node [right] () {EMPTY};
        \end{scope}
    \end{scope}

    \begin{scope}[xshift=13.5cm,yshift=-3cm]
        \begin{scope}[yshift=0cm]
            \draw [draw, thick] (0.5,-0.25) rectangle node [] () {LUT} (3.5,-0.75);
            \draw [thick] (1.25,-0.75) -- +(0,-0.25);
            \draw [thick] (2.0,-0.75) -- +(0,-0.25);
            \draw [thick] (2.75,-0.75) -- +(0,-0.25);
            \draw [thick] (3.75,-0.75) -- (3.75,-0.25) -- (4.0,-0.25) -- (4.0,-0.75) -- (3.75,-0.75) -- (3.875,-0.54) -- (4.0,-0.75);
            \draw [thick] (3.5,-0.5) -- (3.75,-0.5);
            \draw [thick] (4.0,-0.5) -- +(0.2,0.0) node [right] () {$x_4$};
        \end{scope}
        \begin{scope}[yshift=-0.75cm]
            \draw [draw, thick] (0.5,-0.25) rectangle node [] () {LUT} (3.5,-0.75);
            \draw [thick] (1.25,-0.75) -- +(0,-0.25) node [below] () {$x_3$};
            \draw [thick] (2.0,-0.75) -- +(0,-0.25) node [below] () {};
            \draw [thick] (2.75,-0.75) -- +(0,-0.25) node [below] () {};
            \draw [thick] (3.75,-0.75) -- (3.75,-0.25) -- (4.0,-0.25) -- (4.0,-0.75) -- (3.75,-0.75) -- (3.875,-0.54) -- (4.0,-0.75);
            \draw [thick] (3.5,-0.5) -- (3.75,-0.5);
            \draw [thick] (4.0,-0.5) -- +(0.2,0.0) node [right] () {EMPTY};
        \end{scope}
    \end{scope}

    \draw (2.5,1.1) node [] () {\large QUBO Connectivity Data};
    \draw (9,1.1) node [] () {\large Circuit Connectivity Model};
    \draw (16.5,1.1) node [text width=7cm] () {\large Implementation on Fixed Logic};
\end{tikzpicture}
    \end{adjustbox}
    \vspace{-0.5em}
    \caption{Mapping of an original, $4\times4$ in this example, quadratic IM (left) to an equivalent FPGA circuit model preserving QUBO problem connectivity (center), which is then mapped to target FPIA represented by FPGA architecture with $3 \times 2$ CLBs (right). Equivalent nodes must subsequently be shorted by local and global routes, as appropriate. Note that the toy problem presented here is relatively densely connected (50\%), and the tiling technique does not achieve any advantage.
    }
    \label{fig:VPR_model_flow}
\end{figure*}

Many tools and algorithms for implementing circuits on FPGA can be re-purposed to implement FPIA, owing to the similarity in structure between the two. In this study, similarly to \citep{mondalIsingFPGASpintronicsbasedReconfigurable2021}, we leverage the open-source VPR (Versatile Place and Route) tools \citep{betzVPRNewPacking1997,luuVPRFPGACad2009} to pack, place, and route QUBO problems in an FPIA. Specifically, a fictional ``FPGA architecture'' and ``circuit netlist'' are generated to encode, respectively, the capabilities of an FPIA and connection requirements of a QUBO --- the process is visualized for a toy problem in Fig.~\ref{fig:VPR_model_flow}. Each spin is mapped to a flip-flop to ensure a ``circuit'' without combinatorial loops. Each crossbar array row is mapped to an equivalent-width LUT. 

An efficient routing architecture and algorithms require logical equivalence of FPGA CLB's inputs and outputs, i.e., the flexibility of permuting CLB's inputs and outputs. Such a feature is already available in our IMC block because its spins' location and external inputs can be shuffled by appropriately setting up weights in the $(I+O)\times O$ sub-array. To ensure FPIA IMC block's logical equivalence in the VPR tools, the modeled CLB architecture includes a fully-populated crossbar switch with $I+O$ inputs and $O$ outputs connecting CLB's $I$ input pins and its $O$ flip-flop outputs to $O$ $I$-input LUTs. Note that we restrict the VPR packing step to only utilizing $I\times O$ sub-arrays, i.e., CLBs with $O$ $I$-input LUTs in the modeled FPGA architecture, instead of using full $(I+O)\times O$ sub-arrays. This is because the VPR packer cannot be configured to limit the maximum number of external couplings to $I$ (out of available $I+O$ total). With such limitation, CLB's fully-populated crossbar switch functionality is effectively implemented with $O\times O$ portion of sub-array.

Numerous crossbar-compatible memory cells have been explored for IMC applications \citep{Sebastian2020MemoryDA}. In this study, we focus on three representative technologies --- embedded flash (eFlash) with optimistic and pessimistic area scaling models, corresponding to the original \citep{ESF3} and redesigned eFlash \citep{guoCICC17}, respectively, and SRAM \citep{yu16KCurrentBased8T2020}. For the latter, since the memory cell is digital, a differential 2-bit coupling weight, a sufficient precision for the studied benchmark problems, is assumed to be implemented with four SRAM memory cells with 1x- and 2x -width read transistors to encode the bit significance, similarly to  \citep{jaiswal8TSRAMMultibitDPE2019}.

Parameters used for the area and performance modeling are summarized in Table~\ref{tab:area_model}, while die areas of $M\times M$-tile FPIA and baseline architectures implementing $N$-spin IM are estimated as 

\begin{equation}\label{eqn:FPIA_area}
\begin{aligned}
    A_{FPIA} \approx& M^2 \left( I \cdot O \cdot A_{cell} + I A_{wl} + O \left( A_{bl} + A_{sense} \right) \right)\\ & + A_{routing} + \frac{M \cdot I \cdot A_{pewl}}{S_{pe}}  + \frac{M \cdot O \cdot A_{pebl}}{S_{pe}}
\end{aligned}
\end{equation}
\begin{equation}\label{eqn:naive_area}
\begin{aligned}
    A_{baseline} \approx& \left\lceil \frac{N}{S} \right\rceil^{2} \left( S^2 A_{cell} + S A_{wl} + S A_{bl} + A_{ADC} \right)\\ & + \frac{N \cdot A_{pewl}}{S_{pe}} + \frac{N \cdot A_{pebl}}{S_{pe}}
\end{aligned}
\end{equation}

\noindent Here, eFlash and SRAM area assumptions are based on the work in \cite{guoCICC17, ESF3} and \citep{mittalSurveyArchitecturalApproaches2015}, respectively. The selected $S$ value for baseline architecture modeling is optimistic, though achievable with proper bootstrapping to address IR issues \citep{Bavandpour18}. Also very optimistic is $A_{ADC} = 10^6\lambda^2$, based on area-optimized SAR ADCs \citep{kull8bitSARADC2013} that would bottleneck speed if used, though ADC area overhead is also used as a variable parameter in a broader study. Sensing and driving overhead assumptions are similar to \citep{bavandpourACortexEnergyEfficientMultipurpose2020, Bavandpour18}. Programming circuitry follows the design described in \citep{bavandpourACortexEnergyEfficientMultipurpose2020} and is assumed to be shared between IMC blocks, with $A_{pewl}$ / $A_{pebl}$ area overhead required to support a word line or bit line, respectively. We assume, however, that some time-multiplexing of programming circuitry can be exploited without bottle-necking overall operation ($S_{pe}$). Note that we neglect overheads of digital processing and annealing circuitry (that can be packed in the unused corners of a tile), as well as sample-and-hold circuitry (which is much smaller compared to ADC overhead according to \citep{shafieeISAAC2016}) and digital final summation in baseline architecture.  

\begin{table}
    \centering
    \begin{tabular}{lrrr}
        \toprule
        & eFlash optimistic & eFlash pessimistic & SRAM \\
        \midrule
        $A_{fetmin}$ & 50 & 50 & 50 \\
        $A_{cell}$ & 60 & 180 & 600 \\
        $A_{sense}$ & 500 & 2500 & 2500 \\
        $A_{wl}$ & 1500 & 1500 & 1500 \\
        $A_{bl}$ & 1550 & 1550 & 1550 \\
        $A_{pewl}$ & 11300 & 11300 & 1550 \\
        $A_{pebl}$ & 7700 & 7700 & 1550 \\
        $S_{pe}$ & 1000 & 1000 & 1000 \\
        $S$ & 256 & 256 & 256 \\
        \bottomrule
    \end{tabular}
    \caption{Representative parameters (areas in units of $\lambda^2$, the square minimum feature area of the technology) of important reference components of FPIA. }
    \label{tab:area_model}
\end{table}

To explore the trade-off space of FPIA architectures, we search for optimal FPIA tile parameters. To optimize such a process and increase the scale at which we could study these problems, we used SMAC3 \citep{lindaurSMAC3VersatileBayesianOptimization2022}, an implementation of a Sequential Model-Based Optimization \citep{hutterSequentialModelBasedOptimization2011}. In the algorithm configuration variant, random forests are used as the surrogate models, which appears to improve the ability to learn useful trends for loss functions that may return $\infty$ over BO methods with a Gaussian Process surrogate model. We use $R_\textrm{tile}=4$ similar to area-optimal architectures described in Ref. \citep{royOptimizationChannelSegmentation1992}, and use a Wilton topology with each wire segment connected to three other wire segments as described in \citep{betzArchCAD1999_2}, since the results are not sensitive to these parameters. Once an FPIA is packed, placed, and routed, the required array and routing size can be extracted, and we define the figure of merit \emph{Tiling Advantage} as the ratio of baseline (Eq.~\ref{eqn:naive_area}) to FPIA (Eq.~\ref{eqn:FPIA_area}) area.

\section{Modeling Results}

\begin{figure}
    \centering
    \vspace{-0.5em}
        \import{fig/gridsearch_prime82919}{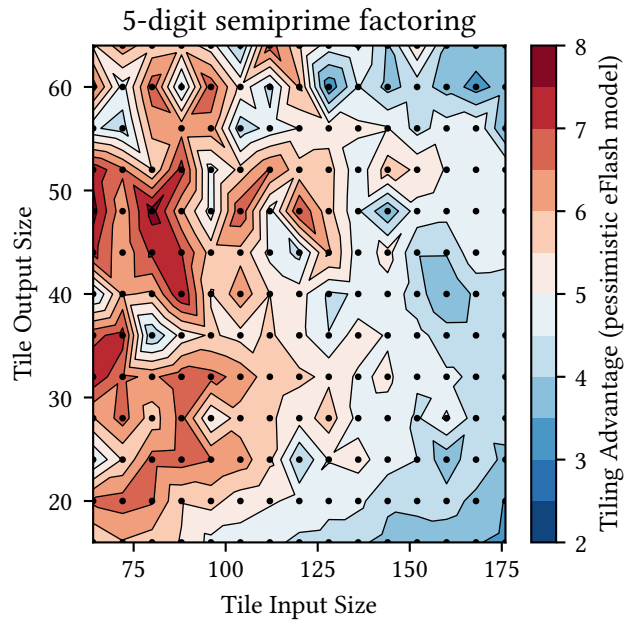}
    \caption{The landscape of tiling area advantage for a 5-digit prime factoring QUBO problem. The input and output connection fractions are fixed for this analysis, $F_I = F_O = 0.2$. 
    }
    \label{fig:prime82919_gridsearch}
\end{figure}

Figure \ref{fig:prime82919_gridsearch} shows a motivation for using Bayesian Optimization. While the tiling advantage is not perfectly monotonic when sweeping $I$ and $O$ parameters, there appears to be a distinct optimal region, e.g., around $I\approx80$ and $O\approx40$ for the shown problem. Further investigation using Bayesian optimization with fixed connection block frequencies ($F_\textrm{I}=0.2$ and $F_\textrm{O}=0.2$) reveals a shared plateau ($O \in [30,50]$, $I \in [80,140]$) of approximate optima for other studied benchmark problems.

\begin{figure}
    \centering
    \import{fig/connection_block_occupancy}{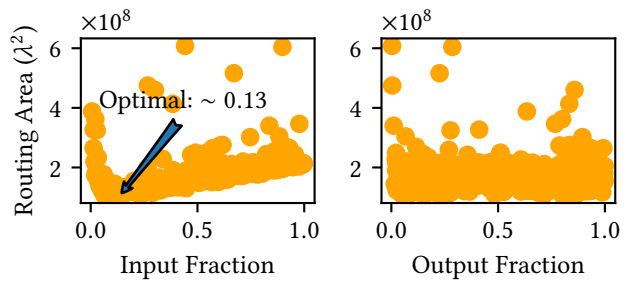}
    \vspace{-0.5em}
    \caption{The impact of $f_\textrm{I}$ and $f_\textrm{O}$ on \emph{routing} area (not including IMC tile area) for 5-digit semiprime factoring problem. Routing area is reciprocal of the tiling advantage since the IMC block area does depend on connection block occupancy.
    }
    \label{fig:marginal_routability_vs_fi_fo}
\end{figure}

We next studied the impact of $F_\textrm{I}$ and $F_\textrm{O}$ on the routing area while fixing $I=80$ and $O=40$, i.e., using quasi-optimal values determined from the previous experiments. The grid search confirms previously chosen $F_\textrm{I} \approx 0.15$ optimal input frequency, close to typically used in FPGA \citep{betzVPRNewPacking1997}, while does not reveal strong patterns in output connection occupancy (Fig. \ref{fig:marginal_routability_vs_fi_fo}). 

\begin{figure}
    \centering
    \import{fig/loc_vs_shr}{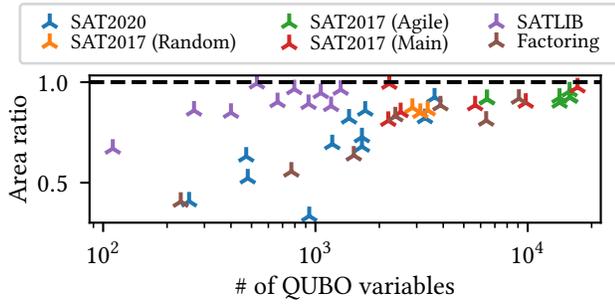}
    \caption{The decrease in area for a shared FPIA architecture (with the same parameters across all mapped problems) as compared to customized FPIA architectures (with parameters optimized per problem) for SRAM implementation.
    }
    \label{fig:loc_vs_shr}
\end{figure}

The practical realization of FPIA hardware implies a fixed architecture that cannot be customized based on the mapped QUBO problem. Therefore, we next focus on``shared'' FPIA architecture based on fixed quasi-optimal $I=140$, $O=40$, $F_\textrm{I}=0.15$,  $F_\textrm{O}=0.2$ determined from previous modeling experiments. Note that the selected $I$ in such shared FPIA must be not smaller than the largest maximum fan-in of the targeted benchmark problems. Naturally, a specifically-optimized architecture is capable of embedding any given problem at least as well as a shared architecture. The relative sub-optimality of the shared embedding is shown in Fig.~\ref{fig:loc_vs_shr}. We note that in the
worst case, a shared FPIA fabric would consume only about 3.4 times more area than an FPIA optimized for that specific problem.

\begin{figure}
    \centering
    \import{fig/pin_underutilization}{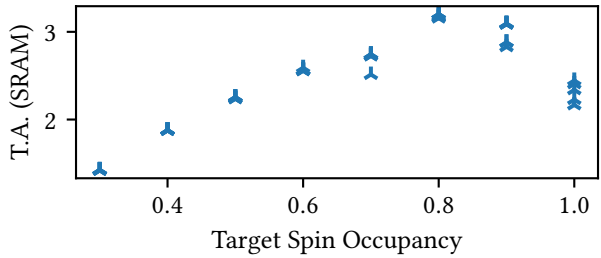}
    \caption{An impact of under-utilizing IMC on tile advantage, studied on the random 3SAT problems for SRAM implementation. 
    }
    \label{fig:underutilization}
\end{figure}

\begin{figure} 
    \centering
    \import{fig/shared_arch_tiling_advantage}{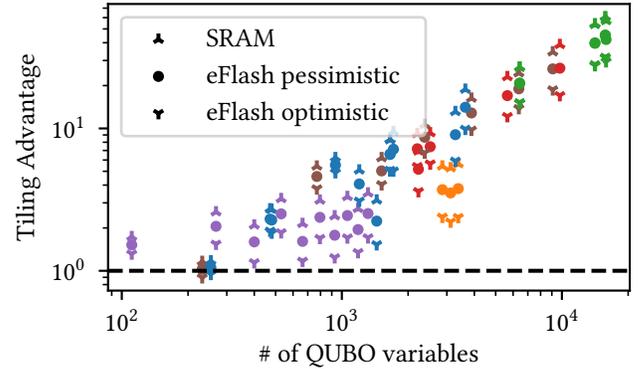}
    \caption{
    Tiling advantage of fixed (``shared'') FPIA architecture for the studied benchmark problems (see Fig. \ref{fig:loc_vs_shr} legend).  
    }
    \label{fig:QUBO_size_vs_tiling_area}
\end{figure}

It has been noted for traditional FPGAs \citep{tessierBalancingLogicUtilization2000} that minimum area is not always achieved for every mapped circuit at the point of maximum logic resource utilization. We observe similar behavior in FPIA by sweeping the target occupancy of each CLB's output pin set, i.e., limiting the number of spins in ICM block to less than $O$. 80\%--90\% occupancy is found to be optimal (Fig.~\ref{fig:underutilization}). Finally, assuming shared architecture, we study tile advantage prospects when exploiting the under-utilization of IMC blocks (Fig. ~\ref{fig:QUBO_size_vs_tiling_area}).

\section{Discussion and Summary}

The modeling results for shared FPIA architecture show that as the problem size increases, so does the Tiling Advantage,  with a slope strongly controlled by the coupling cell area. By design, FPIA architecture can be used to tile smaller problems, but its parameters are not always optimal, and it is often less efficient compared to the baseline approach as smaller problems can fit onto a single physical crossbar of the baseline structure.

The detailed critical path routing delays, i.e., the longest delays of sending a spin value to a distant tile, assuming the optimal circuit parameters for $\lambda=45\si{\nm}$ \citep{KuonAreaDelayFPGA2008,iFAR}, are almost independent of the problem size and always less than 1 ns. IMC latencies are expected to be significantly larger for considered block sizes \citep{bavandpourACortexEnergyEfficientMultipurpose2020}, so that FPIA routing architecture adds negligible performance overhead. Therefore, FPIA is expected to be much faster than baseline architecture that relies on time-multiplexed read-out using shared ADC circuitry.      

\begin{figure}
    \centering
        \import{fig/adc_and_cell_scaling}{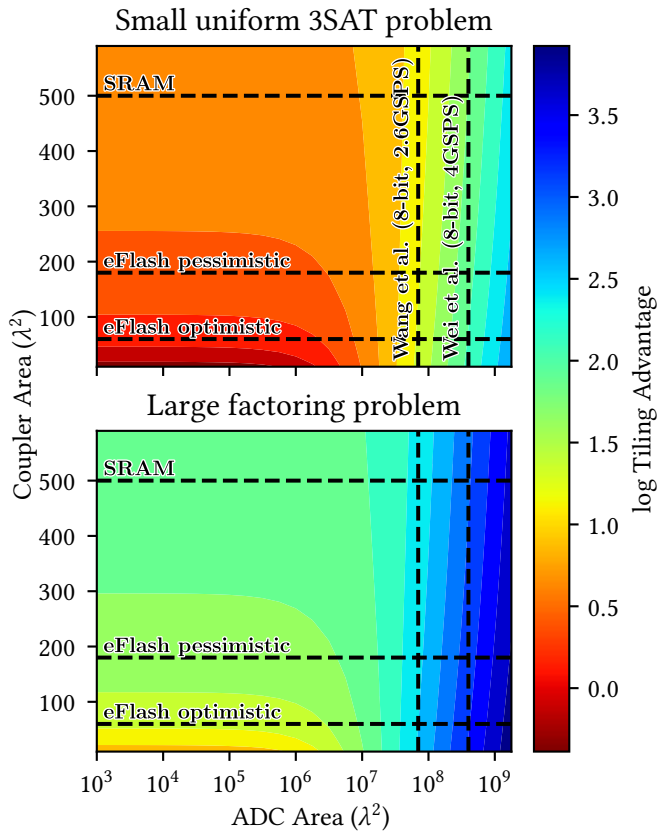}
    \caption{Tile advantage as a function of ADC and IMC unit cell sizes for two representative problems. }
    \label{fig:adc_and_cell_scaling}
\end{figure}

Because of the large contribution of ADC to the baseline design area, we studied the sensitivity of Tiling Advantage to the assumed $A_\textrm{ADC}$. Sweeping $A_\textrm{ADC}$ and memory cell areas (hence indirectly varying practical values of $S$, another critical parameter in our baseline architecture)  reveals a transition from crossbar-dominated to ADC-dominated scaling. (Fig.~\ref{fig:adc_and_cell_scaling}). An approximate expression for the Tiling Advantage is $\approx \frac{N^2 (A_{cell} + \tilde{A}_{ADC})}{N^2 C A_{cell} + A_{routing}}$, where $\tilde{A}_{ADC}$ is the amortized ADC area per cell, and $C$ is the weight utilization improvement factor. Notably, in the ADC-dominated regime, the baseline design area is effectively constant w.r.t. cell area, but the FPIA area is increased, hence decreasing Tiling Advantage. Considering more broader options of ADC designs and memory technologies \citep{wei8Bit4GS120mWADC2013,wangGS8BitTimeInterleavedADC2019} that meet the speed and precision requirements to run the baseline design at least $\SI{10}{\mega\hertz}$, we find that realistic designs would be likely ADC-dominated. 

Future work can improve FPIA directly in the architecture or through co-designing architecture and problem pre-processing. One caveat of shared FPIA architecture is that IMC sub-array minimum horizontal ($I$) dimensions cannot be smaller than the maximum fan-in of any spin of the targeted benchmark problems, which is expected to grow with problem size (Fig.\ref{fig:factoring_weights}b). However, the average fan-in is almost flat (Fig.\ref{fig:factoring_weights}b), and, therefore, a promising algorithmic approach is to break high fan-ins in the original problem by inserting new auxiliary variables. This technique changes the energy landscape of the problem. It hence further requires the characterization of possible changes in the solver's navigational efficiency and the impact on the time-to-solution. Another approach is to utilize heterogeneous IMC blocks containing multiple physical crossbars, which would be chosen during packing to implement either several lower fan-in spins or fewer higher fan-in ones.

FPIA was studied in this work with respect to internally-analog, externally-digital, discrete-time optimization algorithms, but it is generalizable to continuous-time, fully-analog methods. This is because the routing architecture of FPIA requires no time-multiplexing, so wires could be directly used for analog signaling. However, this requires careful simulation of how much delay (and variation in delay) analog solver algorithms can tolerate.

In summary, we have proposed field-programmable architecture for efficiently implementing quadratic Ising Machines and related concepts. Leveraging open-source VPR design automation tools and Bayesian Optimization, we found a set of optimal parameters for the proposed hardware architecture for various practical benchmark combinatorial optimization problems. The modeled improvements are very encouraging, showing up to 60x area and faster operation compared to the baseline approach due to efficient exploitation of the benchmark problem sparsity. 

\ifACM
\begin{acks}
\fi
\ifIEEE
\section{Acknowledgements}
\fi
This material is based upon work supported by the Defense Advanced Research Projects Agency (DARPA) under the Air Force Research Laboratory (AFRL) Agreement No. FA8650-23-3-7313. The views, opinions, and/or findings expressed are those of the authors and should not be interpreted as representing the official views or policies of the Department of Defense or the U.S. Government.
\ifACM
\end{acks}
\fi

\bibliography{IEEEabrv,main}

\end{document}